\documentclass[letterpaper,12pt]{article}
\usepackage{tabularx} 
\usepackage{amsmath}  
\usepackage{graphicx} 
\usepackage{subfig}
\usepackage{caption}
\usepackage[margin=1in,letterpaper]{geometry} 
\usepackage{cite} 
\usepackage{float}

\begin{document}

\title{A Fully Relativistic Treatment of Confined Hydrogen-Like Atoms}
\author{J. M. Noon}
\date{August 13, 2019}
\maketitle

\begin{abstract}
The Dirac equation is used to provide a relativistic calculation of the binding energy of a hydrogen-like atom confined within a penetrable spherical barrier. We take the potential to be Coulombic within the barrier and constant outside the barrier. Binding energies are derived for the ground state of hydrogen for various barrier heights and confining radii. In addition, it is shown that without the introduction of the principle quantum number $n$, all energy states of the confined relativistic hydrogen atom, determined by a single quantum number $k$, transfer into the known energy states of the free relativistic hydrogen atom as the radius of confinement becomes large.
\end{abstract}

\section{Introduction}
The effect of confinement on the energy levels of an atom has been studied by multiple authors, probably first by Michels and deBoer\cite{1}, and then followed by Sommerfeld and Welker who computed the confinement radius at which the binding energy becomes zero.\cite{2} These and more recent authors provided non-relativistic treatments of the problem based on the Schrodinger equation. See in particular Refs. 3-7, the last of which provides additional sources. The intent of this paper is to derive a mathematical model for confined relativistic hydrogen-like atoms, which can then be utilized to investigate the effects of confinement on the relativistic energy levels of hydrogen-like atoms.

The fields of nano-structures and semi-conductor quantum dots have stimulated renewed interest in the problem we consider as a consequence of the need to take account of the effect on atoms from confining boundaries. Because confinement of the atom can cause the energy of its electrons to become relativistic, a relativistic treatment of the problem seems required. However, no previous relativistic treatment of the problem described above has been ascertained.

\section{Theory}
To provide a description of a compressed hydrogen-like atom based on the Dirac equation we assume the wave function of the electron to satisfy the Dirac equation in the form
\begin{equation}
[ -i\hbar c \gamma_{o}\vec{\gamma}\cdot \nabla  +  \gamma_{o}mc^2  + V(r) ] \psi(\textbf{r})  =  E \psi(\textbf{r}),
\end{equation}
where $\gamma_{o}$ and $\gamma$ represent Dirac $\gamma$-matrices, and $m$ and $E$ are the rest mass and total energy of the electron respectively. The solution of Eq. (1) can in general be represented as a four element column matrix dependent on the spherical coordinates, $r$, $\theta$ and $\phi$  of the coordinate vector $\textbf{r}$ as
\begin{equation}
\psi(\textbf{r}) = \frac{{1}}{{r}}\left( \begin{matrix}
G(r)\Omega _{j\ell {m}_j }(\theta ,\phi)  \\
iF(r)\Omega _{j\ell '{m}_j } (\theta ,\phi)
\end{matrix}\right),
\end{equation}
where $\Omega _{j\ell {m}_j } (\theta ,\phi)$ is a two-row spherical spinor and the quantum numbers $\ell$ and $\ell'$ characterize the upper and lower components of the Dirac matrix.

The interest is in the radial coordinate dependence of $ \psi$, expressed through the radial functions $ {G}  {(r)}$ and ${F} {(r)}$, which the Dirac equation connects through the coupled equations
\begin{equation}
\Big[\frac{{d}}{{{dr}}}- \frac{{k}}{{r}}\Big]F(r) +\frac{1}{\hbar c}[E - mc^2 - V(r)  ] G(r) = 0,\tag{3.a}
\end{equation}
\begin{equation}
\Big[\frac{{d}}{{{dr}}}
 + \frac{{k}}{{r}}\Big] G(r) - \frac{{1}}{{{\hbar c}}}
[E + mc^2 - V(r)] F(r) = 0,\tag{3.b}
\end{equation}
where the quantum number $k$ has the two possible values $k = \pm~(j + \frac{{1}}{{2}})$, with $j=l\pm{\frac{1}{2}}$.
Here, for values of the radial coordinate $r$ less than a certain "radius of confinement" $R$, the central potential $V(r)$ is assumed to have the usual Coulomb form
\begin{equation}
V(r) = - Ze^2/r    ~~~~~ (r < R),\tag{4}
\end{equation}
while, for values of $r > R$,  we simulate confinement of the electron by equating $V(r)$ to a constant (barrier potential), denoted $W$
\begin{equation}
V(r) = W  ~~~~~~~~~~~~(r > R).\tag{5}    
\end{equation}

\subsection{Solution of the dirac equation for $\mathbf{r<R}$}
It is useful to introduce the dimensionless coordinate $\rho$  $\equiv$~$2qr$ into Eqs. (3), where we have $q\equiv~\sqrt {(mc^2)^2 - {E}^{2} }/\hbar c$.  Solutions of the equations for $F(\rho)$ and $G(\rho)$ that are finite at the origin can be shown to be expressible in terms of confluent hypergeometric functions of the first kind\cite{8},
\begin{multline}
G(\rho)=A\sqrt{m c^2+E}\rho^{\gamma}e^{-\frac{\rho}{2}}\Big[~_{1}F_{1}\big(\gamma-\frac{Z\alpha E}{\hbar cq}+1,2\gamma+1,\rho\big)+ \\
\Bigg(\frac{k-\frac{Z\alpha m c^2}{\hbar cq}}{\frac{Z\alpha E}{\hbar cq}-\gamma}\Bigg)~_{1}F_{1}\big(\gamma-\frac{Z\alpha E}{\hbar cq},2\gamma+1,\rho\big)~\Big],\tag{6.a}
\end{multline}

\begin{multline}
F(\rho)=A\sqrt{m c^2-E}\rho^{\gamma}e^{-\frac{\rho}{2}}\Big[~_{1}F_{1}\big(\gamma-\frac{Z\alpha E}{\hbar cq}+1,2\gamma+1,\rho\big)- \\
\Bigg(\frac{k-\frac{Z\alpha m c^2}{\hbar cq}}{\frac{Z\alpha E}{\hbar cq}-\gamma}\Bigg)~_{1}F_{1}\big(\gamma-\frac{Z\alpha E}{\hbar cq},2\gamma+1,\rho\big)~\Big],\tag{6.b}   
\end{multline}
where the function $_{1}F_{1}$ has the series representation
\begin{equation}
_{1}F_{1}(a,b;\rho) = 1 + \frac{{a}}{{b}}\rho  +\frac{1}{{2!}}\frac{{{a(a + 1)}}}{{{b(b + 1)}}}\rho^{2}  +  ...\tag{7}
\end{equation}
Here A represents a normalization constant, and we use the notation
\begin{equation}
\alpha  \equiv~e^{2}/\hbar c \\ \tag{8.a}
\end{equation}

\begin{equation}
\gamma \equiv~\sqrt {\gamma ^{2} }  =\sqrt {{k}^{2} {\rm{ -  (}}\alpha {Z)}^{2} }. \tag{8.b}
\end{equation}

\subsection{Solution of the dirac equation for $\mathbf{r>R}$}
We simulate a barrier at $r = R$ by equating the potential $V$ to the constant value $W$ for $r > R$. Introduction of this potential into Eqs. (3.a) and (3.b) converts the equations for the radial functions $G$ and $F$ into the forms
\begin{equation}
\frac{{dG}}{{{d}\rho }} + \frac{k}{\rho}G - \frac{\eta_{1}}{2q} F= 0,\tag{9}
\end{equation}
\begin{equation}
\frac{{dF}}{{{d}\rho }} - \frac{k}{\rho}F - \frac{\eta_{2}}{2q}G = 0,\tag{10}
\end{equation}
where
\begin{equation}
\eta_{1}~ \equiv ~ \left( {\frac{{ mc^2 + E - W}}{{{\hbar c}}}} \right),~~~ \eta_{2}~ \equiv ~ \left( {\frac{{mc^2 -  E + W }}{{{\hbar c}}}} \right).\tag{11}  
\end{equation}
Differentiating Eq. (9) with respect to $\rho$ produces the second order equation for $G$,
\begin{equation}
\frac{{{d}^{2}G }}{{{d}\rho ^2 }} + \frac{k}{\rho}\frac{{dG}}{{{d}\rho}} - \frac{k}{\rho^2 }G
- \frac{\eta_{1}}{2q }\frac{{dF}}{{{d}\rho}} = 0.\tag{12}
\end{equation}
Solving Eq. (9) for $F(\rho)$ and using it in Eq. (10) results in an equation for $\frac{{dF}}{{{d}\rho}}$ in terms of $G$
\begin{equation}
\frac{{dF}}{{{d}\rho}} = \frac{k}{\rho} \frac{2q}{\eta_{1}} \Big( \frac{{dG}}{{{d}\rho}} + \frac{k}{\rho}G \Big)
+  \frac{\eta_{2}}{2q}G,\tag{13}
\end{equation}
the use of which in Eq. (12) gives a second order equation for $G$ that reduces after cancellations to the form
\begin{equation}
\frac{{{d}^{2}G }}{{{d}\rho ^2 }} = \frac{k(k+1)}{\rho^2 }G
+  \frac{\eta_{1}\eta_{2}}{(2q)^2 }G.\tag{14}    
\end{equation}

In order to solve this second order differential equation, it helps to introduce the variable substitution $G(\rho)=\rho \widehat{G}(\rho)$ and to let $\nu=\frac{\sqrt{\eta_{1} \eta_{2}}}{2q} $ in order to produce a new second order differential equation of the form
\begin{equation}
\rho^2\frac{d^2\widehat{G}}{d\rho^2}+ 2\rho\frac{d\widehat{G}}{d\rho}-  [\nu^2\rho^2+k(k+1)]\widehat{G}=0,\tag{15}
\end{equation}
the general solution of which is expressible as
\begin{equation}
\widehat{G}(\rho)=c_{1}\frac{I_{i}(\nu \rho)}{\sqrt{\rho}}+c_{2}\frac{K_{i}(\nu \rho)}{\sqrt{\rho}} ,~~i=\frac{1}{2}\sqrt{(2k+1)^2}.\tag{16}    
\end{equation}

In Eq. (16) above, $I_{i}$ and $K_{i}$ are the modified Bessel functions of the first and second kind of order $i$ respectively. Here, for solutions of Eq. (15) for which $E<mc^2$ and $W<2mc^2$, $\nu$ will remain real and the arguments of the Bessel functions will remain real. Multiplying Eq. (16) by $\rho$ we obtain an equation for $G$,
\begin{equation}
G(\rho)=c_{1}\sqrt{\rho}I_{i}(\nu \rho)+c_{2}\sqrt{\rho}K_{i}(\nu \rho).\tag{17}
\end{equation}
$I_{i}(x)$ increases exponentially as $x$ increases while $K_{i}(x)$ decreases exponentially as $x$ increases.\cite{9} Therefore the requirement that $G$ and $F$ remain finite for all $r > R$ requires that $c_{1}=0$, which reduces Eq. (17) to
\begin{equation}
G(\rho)=c_{2}\sqrt{\rho}K_{i}(\nu \rho).\tag{18}
\end{equation}

From Eq. (9), we see that
\begin{equation}
F(\rho)=\frac{2q}{\eta_{1}}\Big(\frac{dG}{d\rho}+\frac{k}{\rho}G\Big).\tag{19}
\end{equation}
To obtain $\frac{dG}{d\rho}$, we make use of the following recurrence relations from  Ref. 9, namely
\begin{equation}
K_{i-1}-K_{i+1}=\frac{-2i}{x}K_{i},\tag{20.a}
\end{equation}
\begin{equation}
K_{i-1}+K_{i+1}=-2\frac{dK_{i}}{dx},\tag{20.b}
\end{equation}
where the $i$ in the numerator on the right hand side of Eq. (20.a) refers to the subscript $i$. Adding Eq. (20.a) to Eq. (20.b) results in an equation that can be used to determine $\frac{dG}{d\rho}$,
\begin{equation}
\frac{dG}{d\rho}=c_{2}\Big[\frac{1}{\sqrt{\rho}}K_{i}(\nu \rho)\Big(\frac{1}{2}-i\Big)-\nu\sqrt{\rho}K_{i-1}(\nu \rho)\Big].\tag{21}
\end{equation}
Combining Eq. (21) above with equation Eq. (18) and substituting into Eq. (19) results in an equation for $F$:
\begin{equation}
F(\rho)=c_{2}\sqrt{\rho} \Big[\frac{q}{\eta_{1}\rho}K_{i}(\nu \rho)(1-\sqrt{(2k+1)^2}+2k)-\sqrt{\frac{\eta_{2}}{\eta_{1}}}K_{i-1}(\nu \rho)\Big],\tag{22}
\end{equation}
where here $i=\frac{1}{2}\sqrt{(2k+1)^2}$, $\nu=\frac{\sqrt{\eta_{1}\eta_{2}}}{2q}$, and $F(\rho)$ remains finite as $\rho$ approaches infinity as required.

\subsection{Boundary condition at $\mathbf{r=R}$}
The continuity of $ \psi(\mathbf{r})$ required at $r = R$ results in the simultaneous equations
\begin{equation}
G_{r < R}(R) =  G_{r > R}(R),\tag{23.a}
\end{equation}
\begin{equation}
F_{r < R}(R) =  F_{r > R}(R).\tag{23.b}
\end{equation}
After division of Eq. (23.b) by Eq. (23.a), the two equations  are conveniently combined into the "matching equation"\cite{10}
\begin{equation}
\frac{{{F_{r < R}(R)}}} {{{G_{r < R}(R)}}}  =   \frac{{{F_{r >R}(R)}}} {{{G_{r >R}(R)}}},\tag{24}
\end{equation}
independent of the normalization constants $c_{2}$ and $A$. Use of Eqs. (18) and (22) expresses the right hand side of Eq. (24) in the form
\begin{equation}
\frac{1-\sqrt{(2k+1)^2}+2k}{2\eta_{1}R}-\sqrt{\frac{\eta_{2}}{\eta_{1}}}\frac{K_{i-1}(\sqrt{\eta_{1}\eta_{2}}R)}{K_{i}(\sqrt{\eta_{1}\eta_{2}}R)}.\tag{25}
\end{equation}

Meanwhile, the left hand side of Eq. (24) can be evaluated by the use of Eqs. (6.a) and (6.b), in the form

\begin{multline}
\sqrt{\frac{mc^2-E}{mc^2+E}}\times \\ \frac{_{1}F_{1}(\gamma-\frac{Z\alpha E}{\hbar cq}+1,2\gamma+1,2qR)-\Bigg(\frac{k-\frac{Z\alpha m c^2}{\hbar cq}}{\frac{Z\alpha E}{\hbar cq}-\gamma}\Bigg)~_{1}F_{1}(\gamma-
\frac{Z\alpha E}{\hbar   cq},2\gamma+1,2qR)}{_{1}F_{1}(\gamma-\frac{Z\alpha E}{\hbar cq}+1,2\gamma+1,2qR)+\Bigg(\frac{k-\frac{Z\alpha m c^2}{\hbar cq}}{\frac{Z\alpha E}{\hbar cq}-\gamma}\Bigg)~
_{1}F_{1}(\gamma-
\frac{Z\alpha E}{\hbar   cq},2\gamma+1,2qR)}.\tag{26}
\end{multline}
Eq. (24) reduces the computation of the energy of the electron to the solution of an equation for a single unknown, $E$, with parameters $W, k$, $R$, and $Z$.

\section{Results}
Whereas the derived model can be used to compute the effects of confinement on the relativistic energy levels of any hydrogen-like atom, the data presented will be restricted to the hydrogen atom ($Z=1$). In addition, to avoid imaginary arguments in both the Bessel functions and the confluent hypergeometric functions, we restrict total energy $E$ to be less than $mc^2$ and $W$ to be less than $2mc^2$.
\begin{table}[h!]   
\captionsetup{font=footnotesize}
\centering
\begin{tabular}{llllll} \hline 
State & $n$ & $\ell$ & $j$ & $k$ & $E_{bind}$(eV) \\ \hline
$1s_{1/2}$ & 1 & 0 & 1/2 & -1 & -13.606 \\
$2s_{1/2}$ & 2 & 0 & 1/2 & -1 & -3.402 \\
$2p_{1/2}$ & 2 & 1 & 1/2 & +1 & -3.402 \\
$2p_{3/2}$ & 2 & 1 & 3/2 & -2 & -3.401 \\
$3s_{1/2}$ & 3 & 0 & 1/2 & -1 & -1.512 \\
$3p_{1/2}$ & 3 & 1 & 1/2 & +1 & -1.512 \\
$3p_{3/2}$ & 3 & 1 & 3/2 & -2 & -1.512 \\
$4s_{1/2}$ & 4 & 0 & 1/2 & -1 & -0.850 \\
$4p_{1/2}$ & 4 & 1 & 1/2 & +1 & -0.850 \\
$4p_{3/2}$ & 4 & 1 & 3/2 & -2 & -0.850 \\
\end{tabular}
\caption{Electron states in the relativistic free hydrogen atom with their associated binding energies, $n$, $\ell$, $j$, and $k$ values. A more complete table can be found in Ref. 8.}
\label{tab1}
\end{table}

Solutions for the binding energy values of the confined atom, which can be calculated by solving Eq. (24) ($E_{bind} = E - mc^2$), requires a numerical method that avoids taking derivatives. This is because the unknown $E$ exists in the first parameter of the hypergeometric functions, and derivatives with respect to their parameters necessitate further approximation.  This in turn would produce less accurate results for energy. The most convenient numerical method that does not require differentiation is the Bisection method [11], which is what will be employed to help solve for $E$. To check the accuracy of the numerical analysis, Brent's method [12] will also be utilized, which similarly does not warrant differentiation.  

In the case of an unconfined hydrogen atom, the requirement that the solution of the Dirac equation in a coulomb field converges makes it necessary to "cut off" the hypergeometric series. This usually involves setting the first parameter of the confluent hypergeometric series equal to a negative integer, -$n$ ($n>0$), resulting in a description of the state of the atom in terms of the principal quantum number $n$. Here, in the case of the confined atom, no such action is necessary because of the finite value of $R$. Therefore, it would seem that the sate of the electron in this model is distinguished only by the integer $k=\pm{(j+\frac{1}{2})}$, connected to the angular moment quantum number $j$. However, it will be shown later in this section that we still retrieve the usual states of the hydrogen atom.

\begin{figure}[H]
    \captionsetup{font=footnotesize}
    \centering
    \includegraphics[width = .62\textwidth]{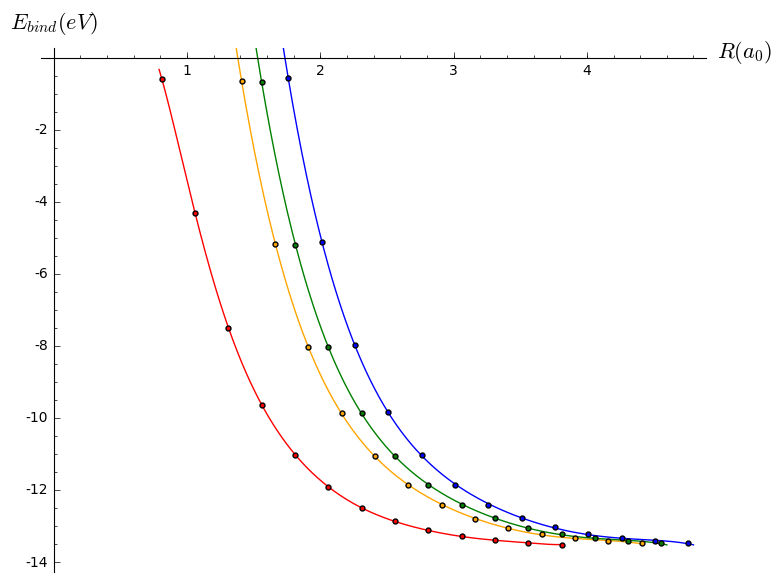}
    \caption{Binding energies for the $1s_{1/2}$ state, or the ground state, of relativistic hydrogen ($k=-1$) as a function of the confining radius $R$. Each curve represents a different barrier height $W$, where $W=0$Ryd, $4$Ryd, $10$Ryd, and $100$Ryd moving from left to right.}
    \label{fig:my_label}
\end{figure}

Figure 1 plots the binding energy as a function of the confinement radius $R$ in units of the Bohr radius $a_{0}$, for multiple fixed values of the barrier height. As the radius of confinement increases, all energy curves approach the value $-13.6$eV associated with the $1s_{1/2}$ state in the free atom. Figure 1 also shows that as the height of the barrier increases, so does the binding energy (given a fixed $R$ value). The relationship between the barrier height and the binding energy can be seen more visibly in Figure 2, where curves are plotted for four different fixed confining radii. To check the accuracy of the results presented thus far, Table 2 compares energy values obtained using Brent's method to energy values obtained using the Bisection method, where each set of energy values were computed using the same domain of $R$ values. The essentially identical results given by both methods serve as a check on the precision of the values displayed in Figures 1 and 2.

\begin{figure}[H]
    \captionsetup{font=footnotesize}
    \centering
    \includegraphics[width = .62\textwidth]{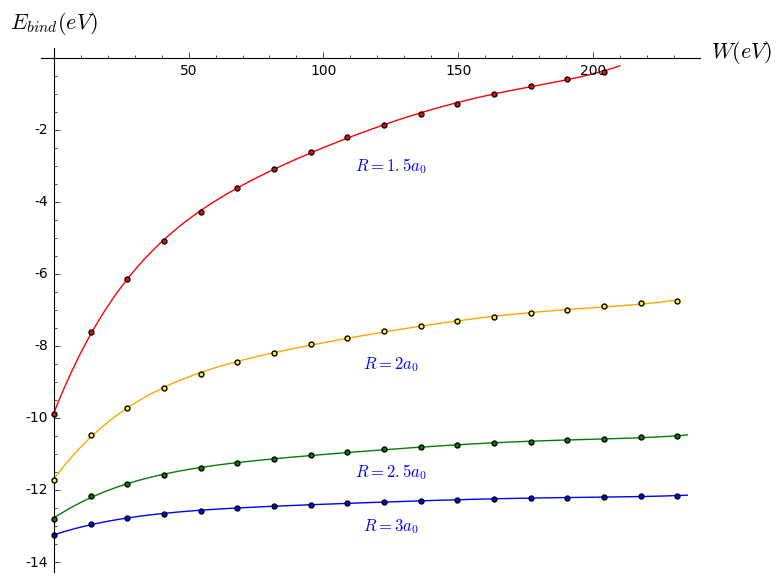}
    \caption{Binding energies for the $1s_{1/2}$ state of relativistic hydrogen as a function of the barrier height at multiple fixed confining radii.}
    \label{fig:my_label}
\end{figure}

\begin{table}[H]   
\captionsetup{font=footnotesize}
\centering
\resizebox{.5\textwidth}{!}{%
\begin{tabular}{lll} \hline 
$R (a_{0})$ & $E (Bisection)$ & $E (Brent's)$ \\ \hline
$0.80$ & $-0.474318624765147$ &
$-0.474318992055487$ \\
$1.10$ & $-4.90025470097316$ & $-4.90025457961019$
\\
$1.40$ & $-8.37346222513588$ & $-8.37346245959634$
\\
$1.70$ & $-10.4843074186938$ & $-10.4843071321375$
\\
$2.00$ & $-11.7343194874702$ & $-11.7343190967804$
\\
$2.30$ & $-12.4782913522213$ & $-12.4782917337725$
\\
$2.60$ & $-12.9248199404683$ & $-12.9248202283052$
\\
$2.90$ & $-13.1944746987429$ & $-13.1944744393113$
\\
$3.20$ & $-13.3578050984070$ & $-13.3578050128417$
\\
$3.50$ & $-13.4567572387168$ & $-13.4567574795801$
\\
$3.80$ & $-13.5165931903175$ & $-13.5165928941569$
\\
\end{tabular}}
\caption{Binding energies for the ground state of relativistic hydrogen ($k=-1$) with barrier height $W=0$Ryd via two numerical methods.}
\label{tab1}
\end{table}

When the barrier wall is very close to the nucleus, the electron can only be in its ground state. Conversely, when the barrier is moved further out from the nucleus, the resulting increase in confining volume accommodates states of the electron with larger orbits, associated with higher values of the quantum numbers $n$ and $\ell$ in the free atom. In particular, Figure 3.(a) graphs the binding energies derived from additional solutions of Eq. (24) with $k=-1$ and larger values of $R$. The graph shows that as $R$ increases, the computed energy values asymptotically approach the binding energies of the free atom in Table 1. corresponding to the electron states $2s_{1/2}$, $3s_{1/2}$, and $4s_{1/2}$, which all have $\ell=0$.

\begin{figure}[H]
    \captionsetup{font=footnotesize}
    \centering
    \subfloat[]{\includegraphics[width=.5\textwidth]{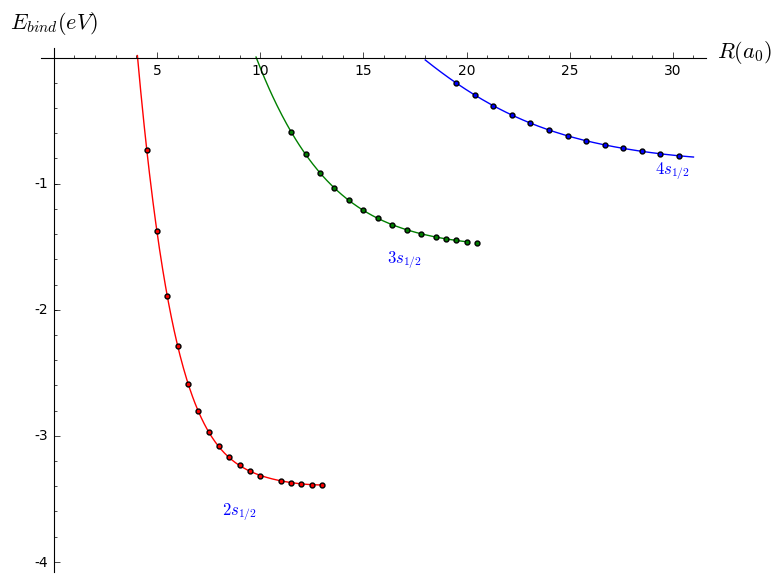}}
    \hfill
    \subfloat[]{\includegraphics[width=.5\textwidth]{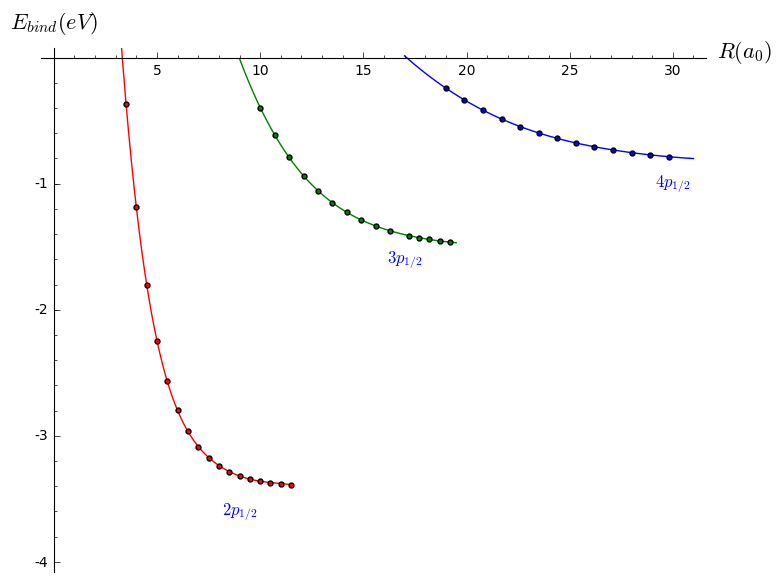}}
    \vspace*{8pt}
    \caption{Binding energies as a function of the confinement radius with $W=0$Ryd. (a) For $k=-1$, the $2s_{1/2}$, $3s_{1/2}$, and $4s_{1/2}$ states appear as $R$ becomes large. (b) For $k=+1$, the $2p_{1/2}$, $3p_{1/2}$, and $4p_{1/2}$ states appear as $R$ becomes large.}
\end{figure}

It is noteworthy that these values are determined only by $k$ and the size of the confining volume restricting the electron to specific states. Figure 3.(b) shows a set of solutions of Eq. (24) for $k=+1$, with the same domain of $R$ values as in Figure 3.(a). Here, as $R$ increases the energy values approach the binding energies of the free atom in Table 1 corresponding to the electron states $2p_{1/2}$, $3p_{1/2}$, and $4p_{1/2}$, all of which have $\ell=1$ and $j=\ell-\frac{1}{2}$.

\section{Conclusion}
Using the Dirac equation, a mathematical model for the confinement of relativistic hydrogen-like atoms in a spherical penetrable barrier was derived. Satisfying the boundary conditions at $r=R$ produced an equation from which (by use of numerical methods) the binding energies for hydrogen-like atoms could be calculated, determined by the parameters $W$, $k$, $R$, and $Z$. Binding energies for the ground state of relativistic hydrogen were calculated as a function of the confining radius at various barrier heights. The binding energies grew with a decrease in the confining radius $R$ and/or a growth in the barrier height, which is a direct consequence of the uncertainty principle. Furthermore, the confined ground state atom's binding energy asymptotically approached the $-13.6$eV binding energy of the free ground state atom as the confining volume grew. Additionally, as the confinement volume grows further (by increasing the value of $R$), excited states of the hydrogen atom (associated with the proper value of $k$) appear without the introduction of the principal quantum number $n$. This implies that the possible energy eigenvalues of the confined system are uniquely determined by the radius of confinement and the value of $k$, as opposed to the energy values of the free atom being determined by $n$ and $k$.

\section{Acknowledgment}
I would like to thank Professor Robert Deck and Professor Jacques Amar for their advice and encouragement throughout this study.

\end{document}